# Thermodynamic properties of strongly coupled plasma in presence of external magnetic field

Mahmuda Begum [a], Deep Kumar Kuri and Nilakshi Das

Department of Physics, Tezpur University, Tezpur-784 028, Assam, India

Electronic mail: mbegumtu@gmail.com

ndas@tezu.ernet.in

Abstract

Thermodynamic properties of a Yukawa system consisting of dust particles in plasma are studied in presence of an external magnetic field. It is assumed that dust particles interact with each other by modified potential in presence of magnetic field. Accordingly, a modified expression for internal energy has been obtained. A molecular dynamics code is developed to calculate this internal energy for the entire system. Based on the values of internal energy given by the code Helmholtz free energy and pressure are calculated for the system. Our study shows novel kind of behaviour for internal energy in presence of magnetic field. Thermodynamic properties are affected significantly by magnetic field. The study helps to express internal energy as a function of Coulomb coupling parameter and magnetic field.

## I. INTRODUCTION

Complex plasma consisting of electrons, ions and charged dust grains usually of micron size provides a rich field for studying some of the fundamental and interesting phenomena of nature like phase transition, interaction mechanism etc. These phenomena become even more complicated in presence of external magnetic field. The presence of magnetic field modifies interaction potential. Yaroshenko et al. [1] have elaborately discussed about various mutual dust-dust interactions in complex plasmas, including the forces due to induced magnetic and electric moments of the grains. Nambu and Nitta [2] have presented a detailed theory of the ``Shukla-Nambu-Salimullah'' (SNS) potential in a magnetized electron-ion plasma, which is anisotropic in comparison with the effective potential in unmagnetized plasma. In dusty plasma phase transition manifested by the appearance of solid BCC and FCC like crystal structure have been observed. For a certain value of screening constant κ, there appears regularized pattern beyond a critical value of coulomb coupling parameter Γ. Thermodynamics of dusty plasma was studied extensively by Hamaguchi et al. [3,4,5]. Totsuji et al. [6] have calculated thermodynamic quantities of a two dimensional Yukawa system. They have found results using giant cluster expansion method in weak coupling regime and by using MD method in strongly coupled regime. Ramazanov et al. [7] have found internal energy and excess pressure of Yukawa system using Langevin dynamics. Presence of magnetic field however alters the internal energy of the particulates in plasma and hence affects the points of phase transition from solid to liquid state. In this work we have calculated free energy and pressure of the system in presence of an external magnetic field and on the basis of this, thermodynamic state of the system is investigated. Molecular dynamics has been widely used for understanding the phenomena of Coulomb crystallization and phase transition in strongly coupled plasma.

Thermodynamic behaviour of dusty plasma strongly depends on the nature of electrostatic interaction. Presence of magnetic field affects the dynamics of plasma particles and this leads to modification of interaction mechanism among dust particles as well as coupling of dust particles with the background plasma. In fact magnetic field may affect the charge accumulated on the grains and forces acting on them.

For the operation of dusty plasma experiments in presence of magnetic field, it is important to identify suitable plasma parameters. A detailed study of thermodynamic properties and phase transition of dusty plasma in presence of magnetic field is important for the point of understanding of fundamental physics as well as for production and control of the properties of dust crystal. It should be possible to see phase transition of dust crystal just by varying the external magnetic field.

Here in this manuscript, we have calculated internal energy of the system by considering the fact that interaction potential among dust grains becomes anisotropic and get modified due to the presence of magnetic field. Then thermodynamic properties like free energy, entropy, pressure etc. are evaluated for a wide range of Coulomb coupling parameter Γ, screening constant κ and for different values of magnetic field. From our study, fitting curve for $u(\kappa, \Gamma)$ is established. This helps in finding values of internal energy for any values of κ and Γ within the limitation of simulation.

The results of our study may give important insight into the physics of crystallization and phase transition in presence of magnetic field. This may find application in various space and laboratory plasma situation.

## II. DESCRIPTION OF THE THEORETICAL MODEL

We consider a 3D dusty plasma of identical, spherical particles of mass $m_d$ and charge $Q_d$ immersed in a neutralizing background plasma subjected to an external magnetic field $\vec{B}$ applied along z-direction. The dynamics of the plasma particles is affected by the magnetic field and as a result the interaction potential among dust particles becomes anisotropic in presence of the field. The interaction potential around a test dust particle with a charge $Q_d$ is [8, 9]

$$\Phi(r) = \frac{Q_d^2}{4\pi\varepsilon_0 r_{ij} f_1} \exp\left(-r_{ij}/\rho_s\right) \qquad (1)$$

where $f_1 = 1 + f$ and $f = (\omega_{pi}^2/\omega_{ci}^2)$, $\omega_{pi}$ and $\omega_{ci}$ beings the ion plasma and ion gyro-frequencies respectively, $\rho_s = \sqrt{f_1}\lambda_{De} \equiv \left(C_S/\omega_{ci}\right)$ is the ion-acoustic gyro-radius, $\lambda_{De}$ is the electron Debye radius $C_S = \lambda_{De}\omega_{pi}$ and is the ion-acoustic speed.

The thermodynamic properties of dusty plasma is characterized by two important parameters: Coulomb coupling parameter Γ and screening constant κ defined as

$$\Gamma = \frac{Z^2 e^2}{4\pi\varepsilon_0 a K_B T_d} \qquad (2)$$

$$\kappa = \frac{a}{\lambda_D} \qquad (3)$$

Where $a = \left(\frac{3}{4\pi n_d}\right)^{1/3}$ is the mean inter-particle distance, $T_d$ is the temperature of the dust grains, $n_d$ is the dust number density. Depending on the values of Γ or κ, dusty plasma may go to strongly coupled regime. The presence of magnetic field affects the shielding of dust particles and this indirectly leads to modification of the two controlling parameters $\Gamma$ and $\kappa$ to $\Gamma_m$ and $\kappa_m$ respectively, where

$$\Gamma_m = (\Gamma/f_1) = \frac{Q_d^2}{4\pi\varepsilon_0 a f_1} \frac{1}{K_B T_d} \quad (4)$$

$$k_m = \left(a/\rho_s\right) = \left(\frac{a}{\sqrt{f_1}\lambda_{De}}\right) \quad (5)$$

### III. THERMODYNAMIC PROPERTIES

In this section we develop a model to investigate thermodynamic properties of strongly coupled dusty plasma column system in presence of an external magnetic field.

The Hamiltonian for a system of N-particles may be written as

$$H = \sum_{j=1}^{N} \frac{|\vec{P}_j|^2}{2m} + U_{ex} \quad (6)$$

where $p_j$ is the momentum of the $j_{th}$ particle and $U_{ex}$ is the excess energy of the dust particles and background plasma particles. Following the procedure of Hamaguchi et al.[3] and using equation (2) as the interaction potential, excess energy $U_{ex}$ may be calculated in presence of magnetic field as

$$U_{ex} = [\frac{1}{2}\sum\sum_{i\neq j}\Phi(\bar{r}_i - \bar{r}_j)] - \frac{NQ_d^2 n}{2\varepsilon_0 K_D^2(1+f_1)} - \frac{NQ_d^2 K_D}{8\pi\varepsilon_0(1+f_1)} + \frac{NQ_d^2}{8\pi\varepsilon_0}\sum_{n\neq 0}\frac{\exp(-K_D L|n|)}{|n|L(1+f_1)}] \quad (7)$$

where $K_D = \frac{1}{\lambda_D}$. In equation(7), the first term on the right-hand side represents the interaction potential between dust grains, the second term represents the free energy of the background plasma that, on average, neutralizes the charge on the particulates. The third term represents the free energy of each sheath, and the fourth term represents the energy of interaction of every particulates and its own images under periodic boundary conditions. In this present problem, the space, mass, time, velocity, energy and external magnetic field strength are normalized by

$\lambda_D$, $m_d$, $\sqrt{\dfrac{m_d \lambda_D^2}{K_B T_d}}$, $\sqrt{\dfrac{1}{\frac{m_d}{K_B T_d}}}$, $K_B T_d$ and $\sqrt{\dfrac{m_d}{4\pi\varepsilon_0 a}}$ .

The dimensionless form of the excess energy per particle in units of $K_B T_d$ may be written as

$$U_{ex} = \dfrac{U}{NK_B T_d} = \Gamma\left[\dfrac{1}{N}\sum_{i\neq j}\sum \dfrac{\kappa}{r'(1+f)}\exp\left(\dfrac{-r'}{\rho_S}\right) - \dfrac{2\pi N r_{av}'}{L'^3(1+f)} - \dfrac{r_{av}'}{2(1+f)} + \dfrac{1}{2}\sum_{i\neq j}\dfrac{\exp(-L'|n|r_{av}')}{|n|L'(1+f)}\right] \quad (8)$$

Here 'N' is the number of dust particles in the simulation box. The thermal component of potential energy is defined as

$$u_{th}(\kappa,\Gamma) = u(\kappa,\Gamma) - u_\alpha(\kappa) \quad (9)$$

Where $u_\alpha(\kappa)$ is the Madelung energy per particle in units of $K_B T_d$. Madelung energy in units of $\Gamma$ can be written as

$$E(\kappa) = \lim_{\Gamma\to\alpha}\dfrac{u(\kappa,\Gamma)}{\Gamma} \quad (10)$$

The magnetic field affects the excess energy term through $f_1$ in equation(7). For a system of 'N' no. of negatively charged dust particles at temperature T in a volume V, Helmholtz free energy per particle in units of $K_B T_d$ be expressed as

$$f = \dfrac{F}{NK_B T_d} \quad (11)$$

and pressure is

$$p = \dfrac{PV}{NK_B T_d} \quad (12)$$

Free energy and pressure can be determined [4] as a function of $(\kappa,\Gamma)$ by using the following relations

$$\dfrac{\partial f}{\partial \Gamma} = \dfrac{u(\kappa,\Gamma)}{\Gamma} \quad (13)$$

$$p = \frac{\kappa}{6}\frac{\partial f}{\partial \kappa} + \frac{\Gamma}{3}\frac{\partial f}{\partial \Gamma} \qquad (14)$$

Further, entropy can be obtained by using the following relation

$$s = -f + \Gamma \frac{\partial f}{\partial \Gamma} \qquad (15)$$

Free energy for fluid and solid state [4] may be calculated by using equation (13) in following way

$$f_{fluid}(\kappa,\Gamma) = \int_0^1 \frac{u(\kappa,\Gamma')}{\Gamma'}d\Gamma' + \int_1^\Gamma \frac{u(\kappa,\Gamma')}{\Gamma'}d\Gamma' + f_{ideal} \qquad (16)$$

Where $\quad f_{ideal} = 3\ln\Gamma + \frac{3}{2}\ln(kT)_{Ry} - 1 + \ln\frac{3\sqrt{\pi}}{4} \qquad (17)$

The total free energy in solid state may be written as

$$f_{solid} = \int_\alpha^\Gamma [u_{th}(\kappa,\Gamma) - \frac{3}{2}]\frac{d\Gamma'}{\Gamma'} + f_{harm}(\kappa,\Gamma) \qquad (18)$$

The harmonic component $f_{harm}$ is calculated [4] using following relation

$$f_{harm} = E(\kappa) + \lim_{N\to\infty}\frac{1}{N}\sum_{k=1}^{3N-3}\ln\frac{\omega_k}{\omega_p} + \frac{9}{2}\ln\Gamma(kT)_{Ry} + \frac{3}{2}\ln\frac{3}{2} \qquad (19)$$

Where $\omega_k$ = Eigen frequencies of an N-particle Yukawa lattice [10]

## IV. DEVELOPMENT OF MOLECULAR DYNAMICS CODE AND STUDY OF PHASE TRANSITION

In order to study the effect of magnetic field on thermodynamic properties and phase transition of Coulomb crystal, a Molecular Dynamics code is developed. The equation of motion for the $i_{th}$ dust particles may be written as

$$m_d \frac{d^2\vec{r}_i}{dt^2} = \vec{F}_i(t)$$

Where $\vec{F}_i(t) = -Q\sum \nabla_i \Phi(r_{ij}) + Q\vec{v}_i(t) \times \vec{B}$ for i= 1, 2, 3,..., N and $j \neq i$. Here, $m_d$ is mass of the dust grain, $r_i$ is the position of the grain i, $F_i$ is the force acting on the $i_{th}$ particle and $\phi_{ij}$ represents *Debye – Hückel* type of interaction potential. For our MD simulation we have taken dust grain mass $m_d = 4.0 \times 10^{-15} Kg$, ion mass $m_i = 1.6726 \times 10^{-27} Kg$, dust density $n_d = 3.74 \times 10^{10} m^{-3}$, ion density $n_i = 1.0 \times 10^{14} m^{-3}$, electron and ion charge $q_e = q_i = 1.602200 \times 10^{-19} C$, electron temperature $T_e = 2323.0 K$ and dust grain radius $r_d = 2.0 \times 10^{-6} m$. The simulation is performed with 686 particles for BCC crystal structure and 500 particles for FCC crystal structure, under periodic boundary conditions, in a 3D cubic simulation box of side L. Each grain is assigned an initial random velocity such that the average kinetic energy corresponds to the chosen temperature $T_d$. Velocity- Verlet algorithm is used to calculate the new position and velocities from the computed forces. The conservation of energy and momentum is verified to check whether the simulation is self consistent, and can be used for new interaction models. The information about structure of dust particles may be obtained from radial distribution function defined as $g(r) = \frac{V}{N} \frac{N(r,\Delta)}{4\pi r^2 \Delta}$. From our simulation g(r) has been computed and plotted across inter-particle distance for different values of magnetic field. Here V is the volume of the simulated region, N is the number of simulated particles, and $N(r,\Delta)$ is the number of particles located in a shell of infinitesimal thickness $\Delta$ from $r - \frac{\Delta}{2}$ and $r + \frac{\Delta}{2}$. Excess energy for the dust particles is calculated by using this simulation.

## V. RESULTS AND DISCUSSION

Based on the expressions as described in section III and using MD simulation, values of excess energy for fluid and solid FCC state is calculated for screening parameter κ=1.4, 2.0, 3.0 and 4.5. Magnetic field is varied in our work over a wide range of values. We have investigated our results for different values of Coulomb coupling parameter Γ starting from fluid to solid regimes. The results are presented in Table 1. Values of Madelung energy E(κ) as defined by equation(8) are given in Table 2 for different values of κ.

**Table1:Excess energy per particle(u\Γ) for different values of Γ and κ with B=0.5Tesla**

| Γ | κ=1.4 | κ=2.0 | κ=3.0 | κ=4.5 |
|---|---|---|---|---|
| 200 | -1.10074639 | -1.01761619 | -1.28880724 | -1.89945038 |
| 240 | -1.10217946 | -1.01854761 | -1.28950521 | -1.90088327 |
| 290 | -1.10381346 | -1.02078936 | -1.29113924 | -1.90251737 |
| 300 | -1.10426951 | -1.02113930 | -1.29159531 | -1.90297347 |
| 305 | -1.10434479 | -1.02118547 | -1.29167055 | -1.90305542 |
| 310 | -1.10433899 | -1.02120882 | -1.29166478 | -1.90321147 |
| 315 | -1.10456050 | -1.02140875 | -1.29188642 | -1.90342698 |
| 320 | -1.10483823 | -1.02170812 | -1.29216426 | -1.90354200 |
| 330 | -1.10556256 | -1.02198237 | -1.29243861 | -1.90381636 |
| 340 | -1.1056892 | -1.02243205 | -1.29288826 | -1.90426611 |
| 350 | -1.10573757 | -1.02260717 | -1.29306343 | -1.90444124 |
| 400 | -1.10692511 | -1.02379476 | -1.29425082 | -1.90562890 |
| 500 | -1.10807425 | -1.0249440 | -1.29540018 | -1.90677802 |
| 600 | -1.10875025 | -1.02605512 | -1.29607968 | -1.90745824 |
| 700 | -1.10933459 | -1.0262043 | -1.29666051 | -1.90803891 |
| 800 | -1.10969932 | -1.02656906 | -1.29702556 | -1.90840319 |
| 900 | -1.1100258 | -1.02689525 | -1.29735168 | -1.90872947 |
| 1000 | -1.11032890 | -1.02719866 | -1.29765176 | -1.90903279 |
| 1200 | -1.11061070 | -1.02748047 | -1.29826549 | -1.90931458 |
| 3000 | -1.11170527 | -1.02821048 | -1.29903108 | -1.91040890 |
| 7000 | -1.11208456 | -1.02895436 | -1.2994104 | -1.91078784 |
| 10000 | -1.11218876 | -1.02905852 | -1.29951495 | -1.91089365 |
| 20000 | -1.11229058 | -1.02916039 | -1.29961619 | -1.91099408 |

**Table2:Madelung energy for FCC Yukawa lattice with B=0.5T**

| κ | $E_{fcc}$ |
|---|---|
| 1.4 | -1.11229058 |
| 2.0 | -1.02916039 |
| 3.0 | -1.29961619 |
| 4.5 | -1.91099408 |

A relation between excess energy and coupling parameter has been obtained from simulation data of Table1 as follows:

$$u(\kappa,\Gamma)=a(\kappa)+b(\kappa)\Gamma+c(\kappa)\Gamma^2+d(\kappa)\Gamma^3 \qquad (20)$$

Where $a(\kappa) = -85.100081 + 178.69097\kappa - 1.72264 \times 10^{-5}\kappa^2 + 10.57784\kappa^3$

$b(\kappa) = -2.46933 + 1.67886\kappa - 0.59475\kappa^2 + 0.05525\kappa^3$

$c(\kappa) = -2.9793 \times 10^{-5} + 4.22371 \times 10^{-5}\kappa - 1.72264 \times 10^{-5}\kappa^2 + 2.08171 \times 10^{-6}\kappa^3$

$d(\kappa) = 1.91671 \times 10^{-9} - 2.47807 \times 10^{-9}\kappa + 9.53688 \times 10^{-10}\kappa^2 - 1.10956 \times 10^{-10}\kappa^3$

We have calculated excess energy for different values of Γ and varying magnetic field B for 0.5T to 0.8T. Screening constant is fixed at κ=3.0. The results are tabulated in Table 3.

**Table3: Excess energy per particle(u\Γ) for different values of B for fixed κ=3.0**

| Γ | B=0.5T | B=0.6T | B=0.8T |
|---|---|---|---|
| 200 | -1.28880724 | -1.43896908 | -1.56885889 |
| 300 | -1.29159531 | -1.44246147 | -1.57168123 |
| 400 | -1.29425082 | -1.44510178 | -1.57315377 |
| 500 | -1.29540018 | -1.44629369 | -1.57445280 |
| 600 | -1.29607968 | -1.44698422 | -1.57483564 |
| 700 | -1.29666051 | -1.44774725 | -1.57571220 |
| 800 | -1.29702556 | -1.44796161 | -1.57628551 |
| 900 | -1.29735168 | -1.44830483 | -1.57654577 |
| 1000 | -1.29765176 | -1.44858123 | -1.57693736 |
| 1200 | -1.29826549 | -1.44489706 | -1.57726225 |
| 1400 | -1.29826549 | -1.4491651 | -1.57745135 |
| 2000 | -1.29866678 | -1.44962680 | -1.57795091 |
| 3000 | -1.29903108 | -1.45003441 | -1.57831092 |

These data are used to find a prediction relation between excess energy and magnetic field for κ=3.0 as follows:

$$u(\Gamma,B) = \alpha_0(\Gamma) + \alpha_1(\Gamma)B + \alpha_2(\Gamma)B^2 + \alpha_3(\Gamma)B^3 \qquad (21)$$

$$\alpha_0(\Gamma) = -12.09078 + 2.01683\Gamma - 4.74916 \times 10^{-5}\Gamma^2 + 9.19298 \times 10^{-9}\Gamma^3$$

$$\alpha_1(\Gamma) = 86:49677 + 12.54264\Gamma + 2.85961 \times 10^{-4}\Gamma^2 - 5.4888 \times 10^{-8}\Gamma^3$$

$$\alpha_2(\Gamma) = -105.04649 + 14.53681\Gamma - 3.41868 \times 10^{-4}\Gamma^2 + 6.56833 \times 10^{-8}\Gamma^3$$

$$\alpha_3(\Gamma) = 41.69137 - 5.68299\Gamma + 1.36264 \times 10^{-4}\Gamma^2 - 2.62071 \times 10^{-8}\Gamma^3$$

**Table4: Free energy for fluid state with B=0.5T for different κ**

| Γ | κ=1.4 | κ=2.0 | κ=3.0 | κ=4.5 |
|---|---|---|---|---|
| 330 | -493.30 | -492.18 | -521.33 | -758.72 |
| 331 | -494.42 | -493.31 | -522.53 | -760.62 |
| 332 | -495.53 | -494.43 | -523.74 | -762.53 |
| 333 | -496.65 | -495.55 | -524.94 | -764.44 |
| 334 | -497.77 | -496.68 | -526.14 | -766.34 |
| 335 | -498.88 | -497.80 | -527.35 | -768.23 |
| 336 | -500 | -498.92 | -528.55 | -770.16 |
| 337 | -501.12 | -500.05 | -529.75 | -772.06 |
| 338 | -502.23 | -501.17 | -530.95 | -773.97 |
| 339 | -503.35 | -502.29 | -532.16 | -775.87 |
| 340 | -504.46 | -503.42 | -533.36 | -777.78 |

**Table5: Free energy for solid FCC lattice with B=0.5T for different κ**

| Γ | κ=1.4 | κ=2.0 | κ=3.0 | κ=4.5 |
|---|---|---|---|---|
| 330 | -493.18 | -491.88 | -521.34 | -771.77 |
| 331 | -494.45 | -493.16 | -522.59 | -772.20 |
| 332 | -495.73 | -494.44 | -523.78 | -772.64 |
| 333 | -497.00 | -495.71 | -524.90 | -773.09 |
| 334 | -498.27 | -496.99 | -526.11 | -773.54 |
| 335 | -499.54 | -498.27 | -527.31 | -774.00 |
| 336 | -500.82 | -499.55 | -528.54 | -774.46 |
| 337 | -502.09 | -500.82 | -529.70 | -774.92 |
| 338 | -503.36 | -502.10 | -530.90 | -775.39 |
| 339 | -504.63 | -503.38 | -532.10 | -775.87 |
| 340 | -505.91 | -504.66 | -533.29 | -776.35 |

### Table6: Values of $\Gamma_T$ for different values of $\kappa$ with B=0.5T

| $\kappa$ | $\Gamma_m$ |
|---|---|
| 1.4 | 331 |
| 2.0 | 332 |
| 3.0 | 336 |
| 4.5 | 339 |

### Table7: Pressure for fluid state with B=0.5T for different values of $\kappa$

| $\Gamma$ | $\kappa$=1.4 | $\kappa$=2.0 | $\kappa$=3.0 | $\kappa$=4.5 |
|---|---|---|---|---|
| 1 | -2.71 | -1.59 | -3.38 | -5.96 |
| 20 | -5.42 | -6.24 | -12.13 | -30.78 |
| 40 | -11.3 | -13.14 | -22.51 | -54.18 |
| 60 | -17.48 | -20.24 | -32.97 | -77.17 |
| 80 | -23.78 | -27.42 | -43.44 | -99.99 |
| 100 | -30.16 | -34.66 | -53.93 | -122.7 |
| 120 | -36.58 | -41.92 | -64.43 | -145.34 |
| 140 | -43.03 | -49.21 | -74.93 | -167.93 |
| 180 | -56.01 | -63.84 | -95.94 | -213 |

The prediction formulae derived in equation (21) is useful to calculate internal energy of the system for any values of $\Gamma$ and magnetic field B. The thermal component of internal energy, $u_{th}$ has been plotted in FIG. 1 and 2 for fluid and solid phases respectively. It is seen clearly from the figures that thermal component of internal energy increases with $\Gamma$ for fluid phase whereas the variation is reverse in the solid regime. Total excess energy has been plotted across $\Gamma$ in FIG. 3(a)-(c) for different values of $\kappa$ and B and in different ranges of $\Gamma$ Analysis of these figures clearly show that internal energy is lowered by the orderly arrangement of the dust particles, internal energy per Coulomb coupling parameter $\Gamma$ has been plotted across magnetic field B [FIG. 4] keeping $\Gamma$ fixed at 617.3 and $\kappa$ at 1.17. Corresponding radial distribution function has been depicted in FIG.5. Lattice correlation factor for the three cases shows that the dusty plasma is at solid crystalline state for B=0.3T, solid-liquid transition point for B=0.7T and in fluid state for B=1.1T. For low magnetic field, internal energy decreases very rapidly until B=0.7T. The turning point in this curve [FIG. 4] indicates the solid-liquid transition point. For further increase in magnetic field, internal energy decreases very slowly. FIG.6 on the other hand shows variation of internal energy for screening parameter $\kappa$. For low values of $\kappa$ variation in $U_{ex}$ is very weak whereas for higher values of $\kappa$, $U_{ex}$ decreases gradually. Higher value of screening parameter corresponds to lower ion thermal motion and this may be the cause of reduction in overall internal energy of the system.

Free energies for fluid and solid FCC states are calculated from excess energy using equations (16) and (18) and the results are presented in Table 4-5. Values of free energies are plotted in FIG.7. From these data we have calculated the value of transition coulomb coupling parameter $\Gamma_T$ for which free energy for solid FCC phase becomes equal to that of the fluid phase of the system. Table 6 gives the value of $\Gamma_T$ for different values of $\kappa$. Free energy is the result of total energy of a system arising due to thermal motion, electrostatic interaction, disorderliness etc. It is the driving force towards equilibrium conditions. A system makes spontaneous transition to a state with lower total free energy. In FIG.7, it is seen that for a fixed value of $\Gamma$. free energy is lower for higher values of $\kappa$. Corresponding value of lattice correlation factor shows that for say $\Gamma=340$, system is in fluid state for $\kappa=3.0$, whereas it is in solid state for $\kappa=3.0$. The system has a tendency to go to fluid state for a particular value of $\Gamma$. Pressure is calculated using equation (14) and the results are presented in Table 7.

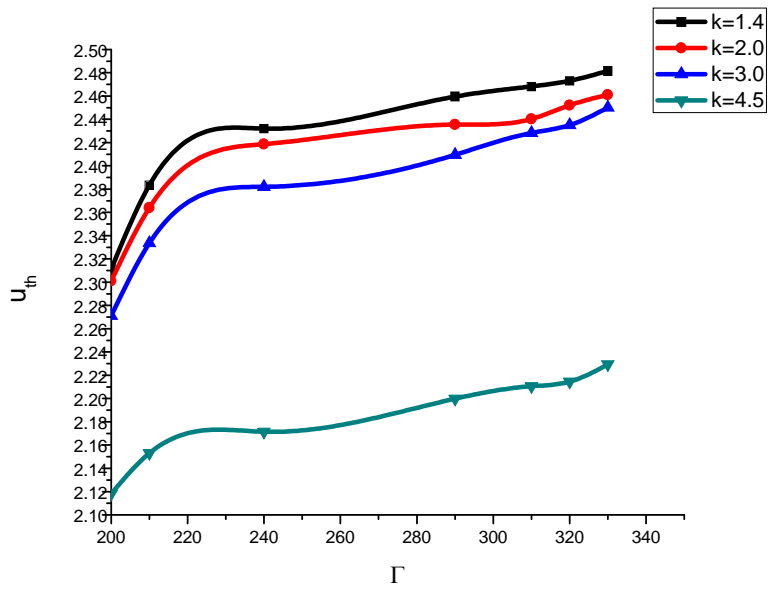

**FIG. 1: Thermal potential energy for fluid phase with B= 0.5 T**

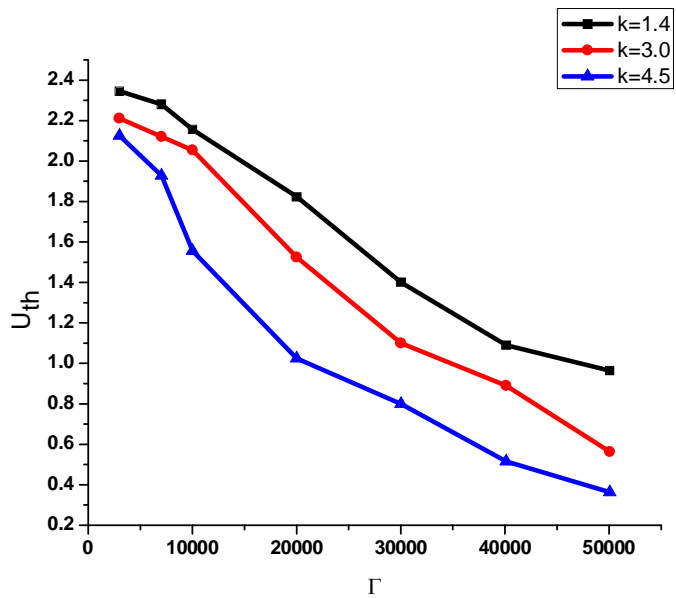

**FIG. 2: Thermal potential energy for solid FCC phase with B= 0.5 T**

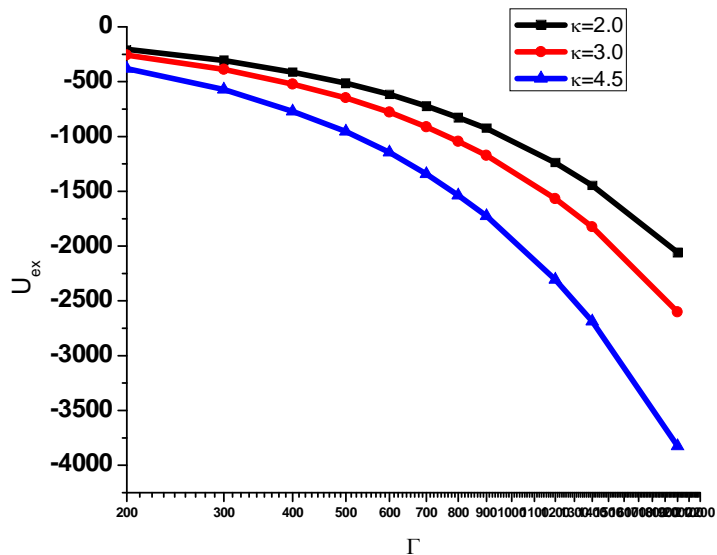

**FIG. 3(a): Excess energy vs. Γ graph for different values of κ with B= 0.5 T**

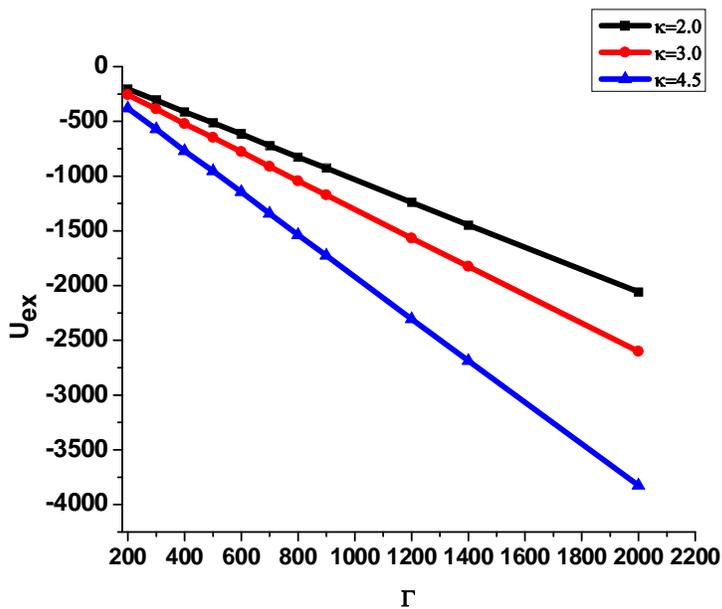

**FIG. 3(b): Excess energy vs. Γ graph for different values of κ with B= 0.5 T**

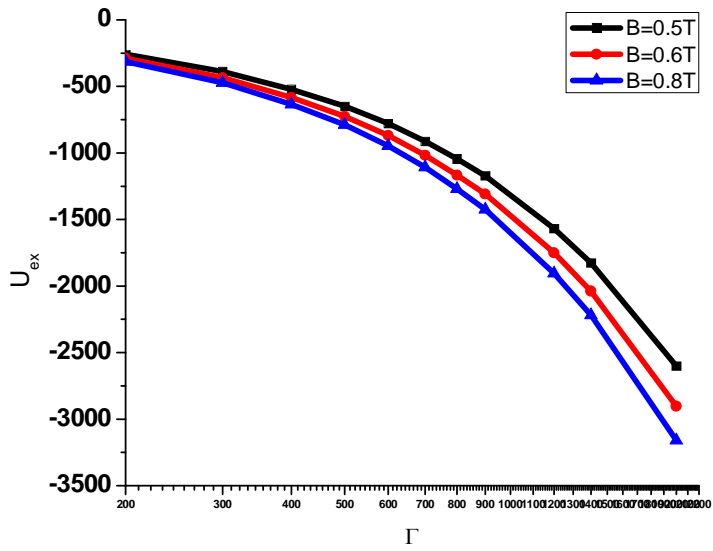

**FIG. 3(c): Excess energy vs. $\Gamma$ graph for different B at $\kappa=3.0$**

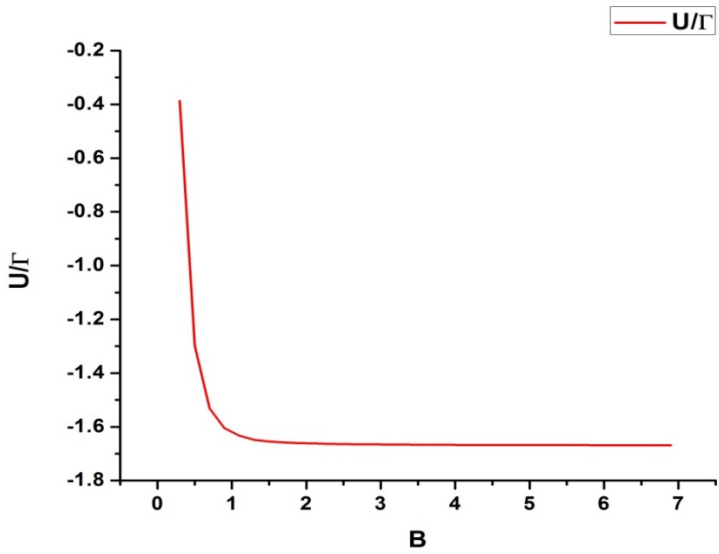

**FIG. 4: U/$\Gamma$ vs. B graph for $\Gamma=617.377$ at $\kappa=1.1774$**

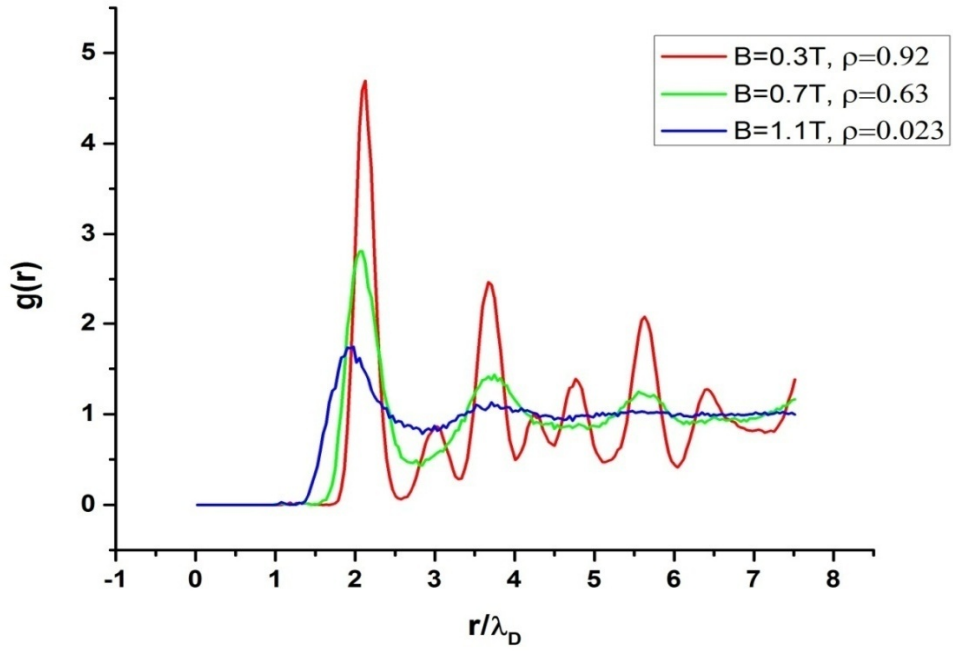

**FIG. 5: Plot of g(r) vs. r/$\lambda_D$ for different B with lattice correlation factor $\rho$ at $\kappa$=1.1774**

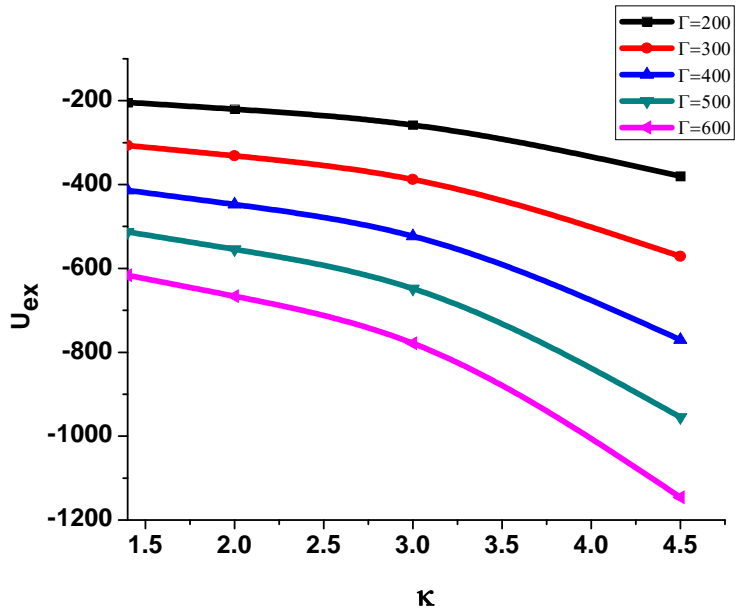

**FIG. 6: Excess energy vs. $\kappa$ graph for different $\Gamma$ at B=0.5T**

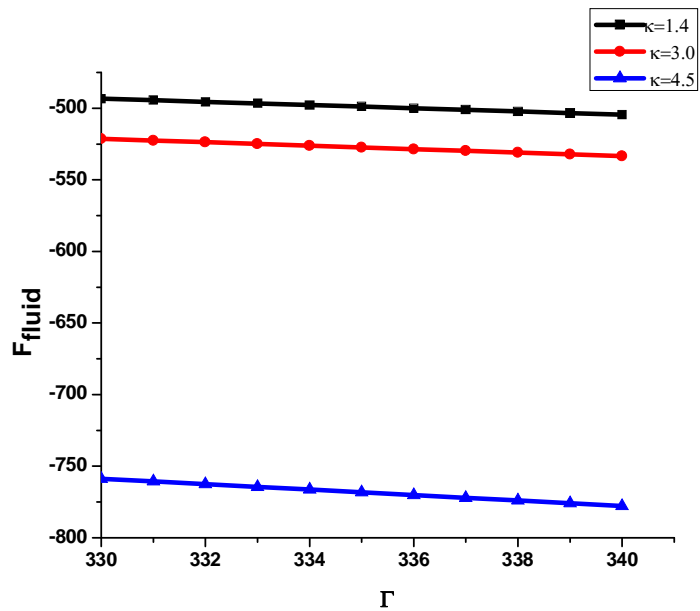

**FIG. 7: Free energy vs.Γ graph for different values of κ**

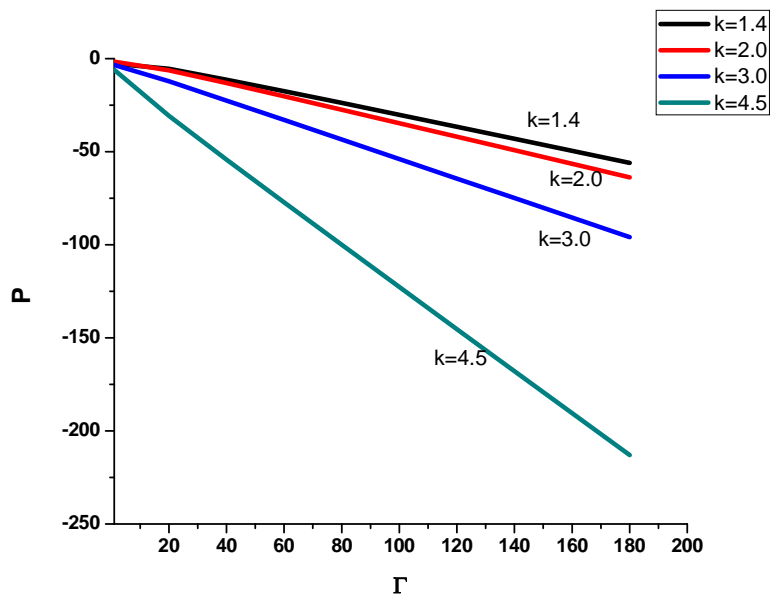

**FIG. 8(a): Excess pressure vs. Γ graph for different values of κ with B= 0.5 T**

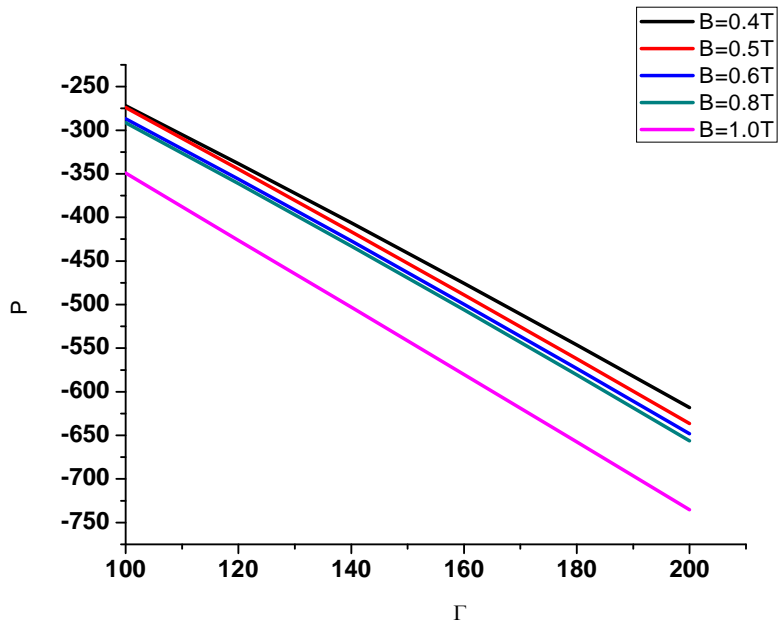

**FIG. 8(b): Excess pressure vs. Γ graph for different B at κ=3.0**

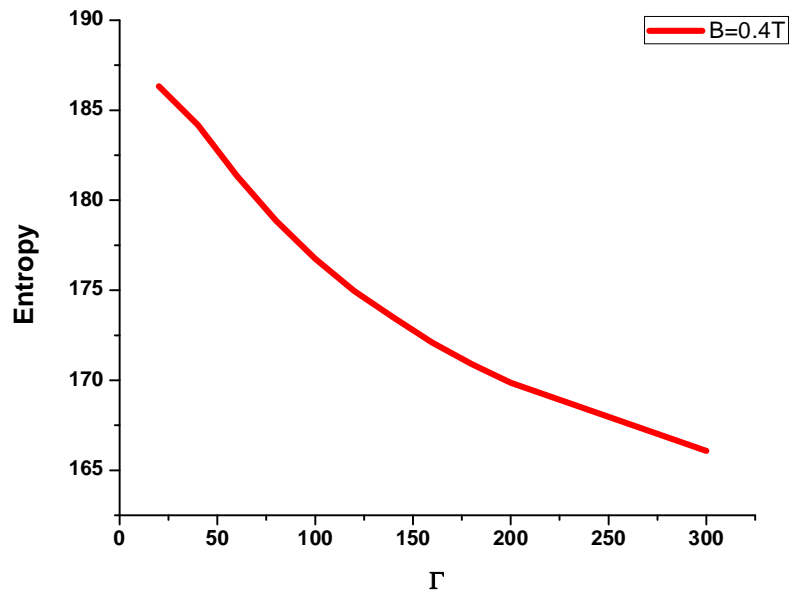

**FIG. 9: Entropy vs. Γ graph with B=0.45T and κ=3.0**

FIG. 8(a)-(b) illustrates excess pressure vs. $\Gamma$ for different $\kappa$, keeping B constant at 0.5T and then for different B, keeping $\kappa$ constant at 3.0. In laboratory dusty plasma the radial electric field that is applied to confine the particles may be responsible for this negative pressure. Finally in FIG.9, entropy has been plotted across $\Gamma$ for $\kappa$=3.0 and B=0.4T. Higher the $\Gamma$, lower in the entropy. This is as expected, since dust particles in crystalline state maintain low degree of disorder under strong Debye-Huckel kind of interaction.

## VI. SUMMARY AND CONCLUSIONS

We have presented an elaborate study on thermodynamics and phase transition of strongly coupled dusty plasma in presence of magnetic field. We have considered that negatively charged dust particles are immersed in plasma and magnetic field is applied along z-direction. The parameters that can control the phase transition may be adjusted in our simulation. We have studied transition from FCC to fluid state by applying external magnetic field. The main results found in this study can be summarized as follows:

(1) A general expression for excess energy has been obtained [equation (7)] for dusty plasma in presence of magnetic field. Molecular Dynamics is used to calculate the values of excess energy for different values of magnetic field. We have obtained a prediction formula for u ($\kappa$,$\Gamma$) as a function of B for any values of coulomb coupling parameter [equation (21)].

(2) The data for excess energy are used to calculate free energy for the system both in fluid and solid FCC regimes, for B=0.4T-1.0T. Free energy is the driving force towards equilibrium conditions. Negative change in free energy seen from Table 4 and 5 indicates that there occurs spontaneous transition to crystalline state. This study helps us to find the value of transition coulomb coupling parameter $\Gamma_T$. It is seen that transition value of $\Gamma$ increases with the increase in screening constant $\kappa$ for fixed magnetic field.

(3) Further, we have calculated pressure and entropy for the system. The plot of excess pressure vs. coulomb coupling parameter shows that pressure is negative for all the values and it becomes more negative as one goes to higher values of $\kappa$.

Entropy is found to decrease with increasing value of $\Gamma$ indicating a well-organized, ordered system.

**The bibliography**